\documentclass[showpacs,preprintnumbers,preprint]{revtex4}
\usepackage{amsmath}
\usepackage{graphicx}
\usepackage{dcolumn}
\usepackage{bm}
\usepackage{amssymb}
\usepackage{makeidx}
\usepackage{amsmath}
\usepackage{lscape}
\usepackage{epsfig}
\newcommand{\ri}{{\mathrm i}}
\newcommand{\p}{\partial}
\newcommand{\bea}{\begin{array}}
\newcommand{\eea}{\end{array}}

\catcode`@=11 \long
\def\@caption#1[#2]#3{\par\addcontentsline{\csname
ext@#1\endcsname}{#1} {\protect\numberline{\csname
the#1\endcsname}{\ignorespaces #2}} \begingroup \small
\@parboxrestore \@makecaption{\csname fnum@#1\endcsname}
{\ignorespaces #3}\par \endgroup} \catcode`@=12

\newcommand{\Q}{\mathbb{Q}}

\newcommand{\la}{\label}

\catcode`@=11 \long
\def\@caption#1[#2]#3{\par\addcontentsline{\csname
ext@#1\endcsname}{#1} {\protect\numberline{\csname
the#1\endcsname}{\ignorespaces #2}} \begingroup \small
\@parboxrestore \@makecaption{\csname fnum@#1\endcsname}
{\ignorespaces #3}\par \endgroup} \catcode`@=12

\begin{document}

\allowdisplaybreaks
  \title { Integrability and supersymmetry of Schr\"odinger-Pauli equations for
neutral particles}
\author{A.G. Nikitin}
\email{nikitin@imath.kiev.ua} \affiliation{ Institute of
Mathematics, National Academy of Sciences of Ukraine,\\ 3
Tereshchenkivs'ka Street, Kyiv-4, Ukraine, 01601}
 \date{\today}

\begin{abstract}
Integrable  quantum mechanical  systems for neutral particles with spin $\frac12$ and nontrivial dipole momentum are classified.  It is demonstrated that such  systems give rise to new
exactly solvable problems of quantum mechanics with clear physical content. Solutions for three of them are given in explicit form. The related symmetry algebras and superalgebras are discussed. The presented classification is restricted to two-dimensional systems which admit matrix integrals of motion linear in momenta.

\end{abstract}
\maketitle
\section{Introduction\label{intro}}

There are two  inspiring notions in quantum mechanics called
supersymmetry and superintegrability. Being formally independent,
both of them are guide signs in searching for exactly solvable
problems. Moreover, some of quantum mechanical systems, like the
Hydrogen atom or isotropic harmonic oscillator,  are both
superintegrable and supersymmetric. Let us note that  just such systems as a
rule are very interesting and important.

A quantum mechanical system with $n$ degrees of freedom is called
superintegrable if it admits more than $n-1$ integrals of motion.
The system is treated as supersymmetric in two cases: when some of
its integrals of motion form a superalgebra, and when its
Hamiltonian has a specific symmetry with respect to the Darboux
transform, called shape invariance \cite{Gen}.

The search for superintegrable systems was started with paper
\cite{wint1} where all the linear and quadratic integrals of motion
for the 2d Schr\"odinger equation with an arbitrary potential had
been presented. The systematic search for  systems whose integrals of motion
are first and second order polynomials in momenta  was performed in
\cite{Mak}, \cite{evan} and \cite{evan2}.

In last decades a number of new important results in this field has
been obtained. In particular,  integrable systems with second
and third order integrals of motion in 2d and 3d Euclidean spaces
had been intensively studied \cite{wint2}-\cite{wint4}, the systems
 both electric and magnetic external fields and
 integrable  systems with spin were discussed in
 \cite{wint5}  and \cite{wint6},
\cite{wint7}, \cite{wint8}.

However, there exist  interesting  integrable systems
which were not studied systematically till now. They are neutral
particles with non-trivial spin and dipole moment (e.g., neutrons)
interacting with an external electromagnetic field. These particles are
described by Schr\"odinger-Pauli equations which include the
Stern-Gerlach or electric dipole terms. A perfect example of such system is the
Pronko-Stroganov model \cite{pron1} describing a neutron coupled to
the field of the constant straight line current. This 2d system
 admits four integrals of motion (including
Hamiltonian) \cite{pron1}, three of which are algebraically
independent. Notice that there exist the relativistic \cite{N1} and
arbitrary spin \cite{pron2} versions of this system which are
integrable and supersymmetric too \cite{N2}.

In addition,  the Pronko-Stroganov system
is shape invariant and can be easily integrated using tools of
supersymmetric quantum mechanics \cite{Vor}, \cite{Gol}. This
circumstance had motivated us to search for other supersymmetric
Schr\"odinger--Pauli equations and classify matrix shape invariant
potentials. In this way a number of new exactly solvable
systems had been found  \cite{NK1}, \cite{NK2}.

In the present paper the integrable 2d
Schr\"odinger-Pauli equations for neutral particles are classified.
Like in paper \cite{wint6} we restrict ourselves to the first order
integrals of motion and classify all external fields which
give rise to integrable and superintegrable systems. As a result a
a new class of integrable systems has been found. The
majority of these systems is supersymmetric also, and one of them is
shape invariant. Thus the germaneness between integrability and
supersymmetry becomes apparent also in quantum mechanical models of
neutral particles.

Integrals of motion of integrable
systems presents powerful tools for finding their exact solutions.
In this paper we restrict ourselves to solving  three of the
obtained systems. One of them appears to be a special case of models
with shape  invariant effective potentials that have recently been
classified  in \cite{NK1} and \cite{NK2}. The other system describes
the neutron interacting with a periodic external field, which have both  discrete and
band energy spectra. One more system is rotationally invariant and includes a superposition of two external fields.

\section{Classification problem\label{DE}}
Let us consider a special class of Schr\"odinger-Pauli equations
describing neutral fermions with non-trivial dipole momentum
interacting with an external field.
The corresponding stationary Schr\"odinger-Pauli equation looks as
follows:
\begin{gather}\la{EP}H\psi({\bf x})=E\psi({\bf x})\end{gather}
where
\begin{gather}\la{Eq}
H=\left(\frac{p^2}{2m}+\frac\lambda{2m}\mbox{\boldmath
$\sigma$}\cdot{\bf B}\right).\end{gather}

 Here $\mbox{\boldmath
$\sigma$} $ is the matrix vector whose components are Pauli
matrices, ${\bf B}={\bf B}({\bf x})$ is a vector of  magnetic field strength, and
vector $\bf x$ represents independent variables. In addition,
$\lambda$ denotes the constant of anomalous coupling which is
usually represented as $\lambda=g\mu_0$ where
 $\mu_0$ is the Bohr magneton and $g$
is the Land\'e factor.

We shall classify integrable systems (\ref{EP}), i.e., find all
Hamiltonians (\ref{Eq}) admitting a sufficient number of integrals
of motion which are linear combinations of momenta with {\it matrix}
coefficients.
In this paper we restrict ourselves to planar systems
depending on two variables $x_1$ and $x_2$. However, bearing in mind possible generalizations of the presented results to 3d systems, we will not exclude the third coordinate a priori.
Till an appropriate moment our analysis will be valid for both 2d and 3d systems.

To obtain more compact formulae let us rescall  variables and reduce
Hamiltonian (\ref{Eq}) to the following form:
\begin{gather} H=-\nabla^2+\mbox{\boldmath $\sigma$}\cdot {\bf B}\label{H}\end{gather}
where $\nabla$ is the gradient vector with componets
$\nabla_a=\frac{\p}{\p x_a}$. To achieve this goal it is sufficient
to change in (\ref{Eq}) $E\to\frac1{2m}E$ and ${\bf B} \to
\frac1\lambda {\bf B}$.

For 3d systems $\nabla^2=\frac{\p^2}{\p x_1^2}+\frac{\p^2}{\p x_2^2}+\frac{\p^2}{\p x_3^2}$ and ${\bf B}$ depends on three variables $x_1, x_2$, and  $x_3$. In the case of 2d systems the third variable $x_3$ and the corresponding derivatives should be deleted. This action can be formalized by imposing the constraint $\nabla_3\psi({\bf x})=0$ and looking for integrals of motion which commute with $H$ and $\nabla_3$.

Let us search for integrals of motion for hamiltonian (\ref{H}) of the following generic
form:
\begin{gather}\label{Q} Q =\sigma^\mu\left(\ri\{\Lambda^{\mu a},\nabla_a\}
+\Omega^\mu\right)\end{gather} where summation is imposed over the
repeated indices $\mu=0,1,2,3$ and $a=1, 2, 3$ or $a=1, 2$ for 3d or 2d systems respectively, $\Lambda^{\mu a}$ and
$\Omega^a$ are functions of ${\bf x}$, $\{\Lambda^{\mu
a},\nabla_a\}=\Lambda^{\mu a}\nabla_a+\nabla_a \Lambda^{\mu a},\
\nabla_a=\frac{\p}{\p x_a}$, $\sigma_\mu$ are Pauli matrices:
\begin{gather*}\label{sigma}\sigma^0=\left(\begin{matrix}1&0\\0&1\end{matrix}
\right),\quad
\sigma^1=\left(\begin{matrix}0&1\\1&0\end{matrix}\right),\quad
\sigma^2=\left(\begin{matrix}0&-\ri\\\ri&0\end{matrix}\right), \quad
\sigma^3=\left(\begin{matrix}1&0\\0&-1\end{matrix}\right).\end{gather*}

By definition, integrals of motion should commute with Hamiltonian,
\begin{gather}\label{com}[H,Q]\equiv HQ-QH=0.\end{gather}

Substituting (\ref{H}) and (\ref{Q}) into (\ref{com}),
using the relations
\begin{gather*}\{\sigma^a,\sigma^b\}=2\delta^{ab},\quad
[\sigma^a,\sigma^b]=2\ri\varepsilon^{abc}\sigma^c, \quad a,b,c
=1,2,3\end{gather*} and equating coefficients for linearly
independent matrices and differential operators, we obtain the
following system of determining equations for coefficients
$\Lambda^{\mu a}$ and $\Omega^\mu$:
\begin{gather}\label{de1}\Lambda^{\mu a}_b+\Lambda^{\mu
b}_a=0,\\\label{de2}\Omega^0_a=0,\\\label{de3}\Lambda^{ab}B^a_b=0,\quad
\Lambda^{0b}B^a_b=\varepsilon^{abc}\Omega^bB^c,\\\label{de4}\Omega^a_b=2
\varepsilon^{acd}\Lambda^{cb}B^d.\end{gather} Here the subindices denote
derivatives w.r.t. the corresponding independent variables, i.e.,
$B^a_b=\frac {\p B^a}{\p x_b}$, etc., and summation is imposed over the
repeated indices.

The system of equations (\ref{de1})-(\ref{de4}) presents the
necessary and sufficient conditions for  commutativity of operators
(\ref{H}) and (\ref{Q}). Equations (\ref{de1}) and (\ref{de2}) are
easy integrated, and their solutions have the following form:
\begin{gather}\Lambda^{\mu a}=C^{\mu[ab]}x_b+C^{\mu a}, \quad
\Omega^0=C^0\label{kil}\end{gather} were $C^{\mu[ab]}, C^{\mu a}$
and $C^0$ are arbitrary constants satisfying the condition
$C^{\mu[ab]}=-C^{\mu[ba]}$. Substituting (\ref{kil}) into (\ref{de3}) and (\ref{de4}) we obtain the overdetermined system of first order partial differential equations for functions $\Omega^a$ and $B^a$.

\section{Determining equations and equivalence transformations for 2d systems}

The determining equations (\ref{de1})--(\ref{de4}) and their partial
solutions (\ref{kil}) are valid for both 2d and 3d equations
(\ref{EP}).
Starting with this point we restrict ourselves to two dimension
systems depending on variables $x_1$ and $x_2$. The corresponding
function $\Lambda^{\mu a}$ in (\ref{kil}) and the related operator (\ref{Q}) are reduced to the
following forms

\begin{gather}\Lambda^{\mu a}=C^{\mu}\varepsilon^{ba}x_b  +C^{\mu
a}, \label{kil2}\\Q=\sigma^\mu \left(C^\mu L+C^{\mu
a}P_a\right)+\sigma^a\Omega^a\la{Q22}
\end{gather} where $C^\mu$ and $C^{\mu a}$ are arbitrary real constants, $a=1,2, \ b=1,2, \ \mu =0, 1, 2, 3,\ \ P_a=-\ri\nabla_a,\ \  L=x_1P_2-x_2P_1$, $\varepsilon^{ab}=-\varepsilon^{ba}$ and $\varepsilon^{12}=1$. Moreover, the number of arbitrary constants in (\ref{kil2}) and (\ref{Q22})
can be reduced using the equivalence transformations which keep the form of Hamiltonian (\ref{H}) up to multiplier $\frac1{\lambda^2}$:
\begin{gather}\label{rot1} x_a\to
x_a+c_a,\\x_a\to R_{ab}x_b,\quad a,b=1,2,\la{rot2}\\
B^k\to \hat R^{kn}B^n,\quad \sigma^k\to \hat R^{kn}\sigma^n,\quad k,n=1,2,3,\la{rot3}\\\label{scal} \ x_a \to \lambda x_a,\ \ B^k\to
\frac{1}{\lambda^2} B^k\end{gather} where $R^1_{ab}$ and $\hat R^{kn}$
are planar and spatial rotation matrices correspondingly, $c_a$ and $\lambda\neq0$
are real constants.

Let $C^0=a\neq0$ then up to shifts (\ref{rot1}) we can set
$C^{01}=C^{02}=0$. Moreover, up to rotation transformations (\ref{rot3}) we can restrict ourselves to $C^3=b,
C^1=C^2=0$. Then, applying  rotations (\ref{rot2}) of  variables $x_1, x_2$ and, if necessary, rotations (\ref{rot3}) with $k,n=1,2$
 we can reduce the remaining
constants $C^{ab}$ to $C^{11}=c_1, C^{22}=c_2, C^{12}=C^{21}=0$
where $c_1$, $c_2$  and $b $ are arbitrary parameters. As a result
we reduce functions $\Lambda^{\mu a}$ in (\ref{kil2}) to the
following form:
\begin{gather}\label{L}\begin{split}&\Lambda^{01}=ax_2, \quad \Lambda^{02}=-ax_1, \quad
\Lambda^{31}=bx_2+d_1, \\& \Lambda^{32}=-bx_1+d_2,\quad
\Lambda^{11}=c_1,\quad \Lambda^{22}=c_2\end{split}\end{gather} while
all the other components of tensors $\Lambda^{\mu a}$ are zeros. In
(\ref{L}) we denote constants $C^{31}$ and $C^{32}$ as $d_1$ and
$d_2$ correspondingly.

 If one of parameters $a$ or $b$ (or both of
them) are zero, we can use (\ref{rot1}) again and reduce the set of functions (\ref{L}) to the
following non-trivial components:
\begin{gather}\label{L2}
\Lambda^{31}=bx_2, \ \ \Lambda^{32}=-bx_1,\ \
 \Lambda^{11}=c_1,\ \
\Lambda^{22}=c_2,\ \ \Lambda^{01}=c_3,\ \ \Lambda^{02}=c_4,\ \
b\neq0\\\label{L3}\Lambda^{01}=ax_2, \quad \Lambda^{02}=-ax_1, \quad
\Lambda^{11}=c_1,\quad \Lambda^{22}=c_2,\quad a\neq0,\\\label{L4}
\Lambda^{11}=c_1,\quad \Lambda^{22}=c_2,\quad \Lambda^{01}=c_3,\quad
\Lambda^{02}=c_4.\end{gather} The corresponding symmetry operators
(\ref{Q22}) look as follows:
\begin{gather}\label{Q1}\begin{split}&Q=aL+\sigma_3b\hat L+\sigma_1c_1P_1+
\sigma_2c_2P_2+\sigma_a\Omega_a,\quad
ab\neq0,\end{split}\\\begin{split}&Q=\sigma_3b
L+(c_3+\sigma_1c_1)P_1+ (c_4+\sigma_2c_2)P_2+\sigma_a\Omega_a,\quad
b\neq0,\end{split}\label{Q2}\\Q=a L+\sigma_1c_1P_1+
\sigma_2c_2P_2+\sigma_a\Omega_a,\quad
a\neq0,\label{Q3}\\Q=(c_3+\sigma_1c_1)P_1+
(c_4+\sigma_2c_2)P_2+\sigma_a\Omega_a\label{Q4}\end{gather} where $
\hat L={\hat x}_1\hat P_2-\hat x_2\hat P_1,\quad \hat x_1=
x_1-\frac{d_2}b,\quad \hat x_2= x_2+\frac{d_1}b,\quad \hat
P_a=-\ri\frac{\p}{\p \hat x_a}.$ In addition, without loss of
generality we can set $a=1$ in (\ref{Q1}) and (\ref{Q3}) and $b=1$
in (\ref{Q2}).

 Notice that for $c_1=c_2=c_5=0$ operators
(\ref{Q3}) and (\ref{Q4}) are reduced to the following form:
\begin{gather}\label{Q51}Q= aL-\frac{k}2\sigma_3,\\Q=c_3P_1+
c_4P_2- \frac{n}2\sigma_3,\quad c_3c_4=0\label{Q61}\end{gather} where $k,\ n, c_3$
and $c_4$ are real parameters.
Indeed, in accordance with (\ref{de4}), in this case functions
$\Omega^{a}$ are reduced to constants, and hermitian matrix
$\sigma_a\Omega^a$ is diagonalizable. Just operators (\ref{Q51}) and
(\ref{Q61}) represent Lie symmetries which could be admitted by
equation (\ref{EP}).

 Equations (\ref{Q1})--(\ref{Q61}) give
representatives of the family of operators  (\ref{Q22})
defined up to equivalence transformations (\ref{rot1})--(\ref{scal}). In order these operators to be integrals of motion for
Hamiltonian (\ref{H}) functions $B^1, B^2, B^3$ and $\Omega^1,
\Omega^3,\Omega^3$ have to satisfy equations (\ref{de3}) and
(\ref{de4}) which, in view of (\ref{L})--(\ref{L4}), take the
following form:
\begin{gather}\begin{split}&
\Omega^3_1=-2c_1B^2,\quad \Omega^3_2=2c_2B^1,\\&
\Omega^1_1=-2(b x_2+d_1)B^3,\quad \Omega^2_2=2(d_2-b
x_1)B^1,\\& \Omega^2_1=2c_1B^3+2(b x_2+d_1)B^1,\\&
\Omega^1_2=-2c_2B^3+2(b\hat x_1-d_2)B^1,\\&
b\left(\hat x_1B^3_2-\hat x_2B^3_1\right)
=c_1B^1_1+c_2B^2_2,\\&
a(x_1B^1_2-x_2B^1_1)+\Omega^3B^2-\Omega^2B^3=c_3B^1_1+c_4B^1_2,\\&
a(x_1B^2_2-x_2B^2_1)+\Omega^1B^3-\Omega^3B^1=c_3B^2_1+c_4B^2_2,\\&
a(x_1B^3_2-x_2B^3_1)+\Omega^2B^1-\Omega^1B^2=c_3B^3_1+c_4B^3_2.\end{split}
\label{e1}
\end{gather}
where parameters $d_1,\ d_2,\ c_3,\ c_4,\ a$ and $b$ satisfy the
conditions $ac_3=ac_4=0; \ d_1=d_2=0\ \text{if}\ ab=0\ \texttt{and}\
a^2+b^2>0$.

Thus the problem of classification of superintegrable planar
Schr\"odinger-Pauli equations admitting first order integrals of
motion is reduced to finding the general solution of determining
equations (\ref{e1}). These equations should be solved consequently for all sets of parameters $a, b, d_1, d_2, c_1, c_2,
c_3, c_4$ present in operators
(\ref{Q1})--(\ref{Q61}).  Thus for operator of type (\ref{Q1}) we set in (\ref{e1}) $c_3=c_4=0,\ a\neq0,\ b\neq0$ while the remaining parameters are arbitrary, etc.
\section{Lie symmetries}
Let us start with an important class of symmetry operators which are
generators of continuous groups. They are the   special cases of
operators (\ref{Q22}) corresponding to $\Lambda^{\mu a}=0$ for all
$\mu\neq0$. Up to equivalence all such operators  are given by equations (\ref{Q51}) and (\ref{Q61}). To classify the external fields ${\bf B}=(B_1, B_2, B_3)$ such that Hamiltonian (\ref{H}) commutes with operators (\ref{Q51}) and (\ref{Q61}) it is sufficient to solve equations (\ref{e1}) with $\Omega_1=\Omega_2=b=d_1=d_2=0, \Omega_3=\frac{n}2$ or $\Omega_3=\frac{k}2$, while parameters $c_3, c_4, n, k$ and $a$  should satisfy the conditions $a(c_3^2+c_4^2)=0$ and $nk=0$. As a result we obtain the list of the external fields and the corresponding symmetries presented in Table 1. A constant external field is not presented here since the corresponding Hamiltonian can be reduced to a direct sum of two Hamiltonians with trivial interaction terms.
\allowdisplaybreaks
\begin{center}
 \bf Table 1. External fields with Lie symmetries
 \end{center}

 \vspace{2mm}

\begin{tabular}{|c|c|c|c|}
\hline
No&
 {External field }& {Symmetry operators}&Comments\\
\hline
1.&$\begin{array}{c}B^1=\cos(k\theta)f_1(r)+\sin(k\theta)f_2(r),\\
B^2=\cos(k\theta)f_2(r)-\sin(k\theta)f_1(r),\\
B^3=f_3(r)\end{array}$&$\tilde Q_1= L+\frac{k}2\sigma_3$
&$\nabla\cdot {\bf B}\neq0$\\
\hline
2.&$\begin{array}{c}B^1=\mu{\cos(k\theta)}r^{-k},\ \
B^2=\mu{\sin(k\theta)}r^{-k},\\
B^3=f_3(r)\end{array}$&$\tilde Q_1$& \\
\hline
3.&$\begin{array}{c}B^1=\mu{\cos(\theta)}f_1(r),\ \
B^2=\mu{\sin(\theta)}f_1(r),\\
B^3=f_2(r)\end{array}$&$Q_1=L+\frac{1}2\sigma_3$& \\
\hline
4.&$\begin{array}{c}B^1=\cos(\delta x_1)f_1(x_2)+
\sin(\delta x_1)f_2(x_2),\\ B^2=\cos(\delta
x_1)f_2(x_2)-\sin(\delta x_1)f_1(x_2),\\
B^3=f_3(x_2)\end{array}$&$\tilde Q_2=P_1
- \frac{\delta}2\sigma_3$&$\nabla\cdot {\bf B}\neq0$\\
\hline
5.&$\begin{array}{c}B^1=\mu\exp(-x_2)\cos x_1,
\\ B^2=-\mu\exp(-x_2)\sin x_1, \ \
B^3=f_3(x_2)\end{array}$&$Q_2=P_1
- \frac{1}2\sigma_3$&$\bea{c}\text{Eq. (\ref{Ha}) is}\\\text{shape invariant}\eea$\\
\hline
6.&$B^1=B^2=0,\quad B^3=f({\bf x})$&$\sigma_3$&decoupled \\
\hline
7.&$B^1=B^2=0,\
B^3=f(x_1)$&$P_2,\quad \sigma_3$&decoupled\\
\hline
8.&$B^1=B^2=0,\
B^3=f(r)$&$L,\ \sigma_3$&decoupled\\
\hline
\end{tabular}

\vspace{2mm}

Here \begin{gather}\label{pc}r=\sqrt{x_1^2+x_2^2},\quad
\theta=\arctan\frac{x_2}{x_1},\end{gather}
$f_1(\cdot),\ f_2(\cdot),\ f_3(\cdot)$ and
$f(\cdot)$ are arbitrary functions, and  $\delta=0$ or $\delta=1$.

For B given in Items 1 and  4 to be single valued and  $\tilde Q_1$ to generate finite
rotations for arbitrary values of theta, parameter $k$ must be an integer.

The fields $\bf B$ presented in Items 2 and 3 of Table 1 are  particular cases of the field presented in Item 1. We specify these particular cases since the corresponding vector $\bf B$ is divergent-free, i.e., satisfies the condition
\begin{gather}\label{div}\nabla\cdot{\bf B}=0\end{gather} which is not valid for more general fields given in Item 1. This condition is necessary if we interpret $\bf B$ as a vector of magnetic field strength. Analogously,  the divergent-free external field presented in Item 5 is a particular case of the more general field given in Item 4.

In the cases enumerated in Items 6 -- 8 Hamiltonian (\ref{H}) is reduced to a direct sum of Hamiltonians with scalar potentials, and the corresponding equation (\ref{EP}) is reduced to the system of two  decoupled equations. This fact is indicated in the fourth column of Table 1.

Thus we find  Hamiltonians
(\ref{H}) whose integrals of motion are generators of Lie
symmetries. One more external field for which (\ref{H}) admits Lie symmetries (but also a more general symmetry equivalent to (\ref{Q4}) with $c_1=c_2=c_4=0, c_3\neq0$) is presented in Item 1 of Table 2.

Let us note that all obtained Hamiltonians  admit a straightforward
generalization which keeps its symmetries. Namely, in addition to
(\ref{H}), we can consider a more general Hamiltonian
\begin{gather}\label{Ha} H=-\nabla^2+\mbox{\boldmath $\sigma$}\cdot{\bf B}
+\omega{\bf B}^2\end{gather} where $\omega$ is an additional
coupling constant.

Setting in (\ref{Ha}) $\omega=0$ we obviously come to the previous Hamiltonians (\ref{H}). However, for all external fields presented in Table 1 the
generalized Hamiltonians (\ref{Ha}) admit the same Lie
symmetries as Hamiltonian (\ref{H}).

One more generalization of Hamiltonian (\ref{H}) which keeps its symmetries can be written as:
\begin{gather}\label{HaHa} H=-\nabla^2+\mbox{\boldmath $\sigma$}\cdot{\bf B}
+ V\end{gather} where $V=V(r)$ and $V=V(x_2)$ for the cases presented in Items 1, 2, 8 and 3, 4, 7
correspondingly.

\section{Non-Lie integrals of motion} In this section we
present integrals of  motion which are not
generators of Lie symmetries. Doing this we restrict ourselves to
the first order integrals of motion whose general form is given by
equation (\ref{Q}) where at least one of coefficients
$\Lambda^{ab},\ a=1, 2, 3$, is nonzero. Such integrals of motion can be represented in the forms given by equations (\ref{Q1})--(\ref{Q4}). Moreover, the latest equation should be supplemented by the condition $c_1^2+c_2^2\neq 0$.

To evaluate the corresponding external fields $\bf B$ and functions $\Omega^a$ it is necessary to solve equations (\ref{e1}) where parameters $a, b, c_1, c_2, c_3, c_4, d_1, d_2$ should satisfy one of the following sets of conditions:
\begin{gather}ab\neq0,\quad c_3=c_4=0,\la{co1}\\a=0,\quad b\neq0,\quad d_1=d_2=0,\la{co2}\\a\neq0,\quad b=0, \quad c_3=c_4=d_1=d_2=0,\la{co3}\\a=b=0, \quad c_1^2+c_2^2\neq0.\la{co4}\end{gather}
Moreover, conditions (\ref{co1}), (\ref{co2}), (\ref{co3}) and (\ref{co4}) correspond to symmetries (\ref{Q1}), (\ref{co2}), (\ref{Q3}) and (\ref{Q4}) respectively. In addition, equivalence transformations (\ref{rot3}) can be used to simplify the form of $Q$ with fixed $\Lambda^{\mu a}$ and $\Omega^a$.

Solving equations (\ref{e1}) under conditions (\ref{co1})--(\ref{co4}) we obtain the list of external fields and the related integrals of motion which is presented in Table 2.

\begin{center}
 \bf Table 2. External fields and higher symmetries
 \end{center}

 \vspace{2mm}

\begin{tabular}{|c|c|c|c|}
\hline
 No&{External field }& {Symmetry operators}&$\bea{c}\text{Comments}\eea$
 \\
\hline
1.&$\begin{array}{c}B^1=\mu\cos x_1,\
B^2=\mu\sin x_1,\\ B^3=\nu\end{array}$&$\begin{array}{c}Q_2= P_1-\frac{1}2\sigma_3, \ P_2,\\Q_3
=\sigma_3(P_1-{\nu})\\-{\mu}(\sigma_1\cos x_1+\sigma_2\sin x_1)\end{array}$&$\bea{c}\text{supersymmetric,}\\\text{see (\ref{qr}), (\ref{qr2})}\eea$\\
\hline
2.&$\begin{array}{c}
 B^1=\frac{{\mu}k\sin(k\theta)}{r^2}, \\
B^2=-\frac{{\mu k}\cos(k\theta)}{r^2},
 \ B^3=\frac{k\nu}{r^2}\end{array}$& $\begin{array}{c}\tilde Q_1=L+\frac{k}2\sigma_3,\\
Q_4=\sigma_3(\tilde Q_1+\nu)\\-\mu(\sigma_1\sin(k\theta)-\sigma_2\cos(k\theta))
\end{array}$&$\bea{c}\text{conformal,}\ \text{see (\ref{AL})};\\\text{single valued}\\\text {for integer} k \eea$\\
\hline
3.&$\begin{array}{c}B^1= \frac{\mu^2x_2}{2\sqrt{\nu^2-\mu^2r^2}},\\
 B^2= -\frac{\mu^2x_1}{2\sqrt{\nu^2-\mu^2r^2}},\
B^3=\frac\mu2\end{array}$&$\begin{array}{c}Q_1=L+\frac{1}2\sigma_3,\\
Q_5=\sigma_1P_1+\sigma_2P_2-\frac\mu2(\sigma_1x_2\\-
\sigma_2x_1)-\frac12\sigma_3\sqrt{\nu^2-\mu^2r^2}\end{array}$&$\bea{c}
\text{supersymmetric,}\\\text{see (\ref{SA3}), (\ref{SA31})}\eea$\\
\hline
4.&$\begin{array}{c}B^1= \frac{x_2\varphi'}{r},\\
B^2=- \frac{x_1\varphi'}{r}, \
B^3=-{\mu(r\varphi)'}\end{array}$&$\begin{array}{c}Q_1,
\\
Q_6=\sigma_1P_1+\sigma_2P_2+\mu(\sigma_3Q_1\\+\sigma_1x_2\varphi-
\sigma_2x_1\varphi)+\sigma_3\left(\varphi+\nu\right)\end{array}$&
$\bea{c}\text{supersymmetric,}\\\text{see (\ref{SA1}),
(\ref{SA2}), (\ref{SA11})}\eea$\\
\hline
\end{tabular}

\vspace{2mm}
Here  $\varphi=\varphi(r)$ is a solution of the following algebraic
equation
\begin{gather}\label{MF22}\left(\mu^2r^2+1\right)\varphi^2+2\nu\varphi=
c, \end{gather} $\mu,\ \nu$,  $k$ and $c$ are real parameters. In
particular, for $\nu=0$, $c=\omega^2$ and $c=0,\ \nu=-4\omega$ we
obtain the following versions of the field presented in Item 4:
\begin{gather*}B^1=-\frac{\omega\sin\theta\sinh\rho}{\cosh^3\rho},
\quad B^2=\frac{\omega\cos\theta\sinh\rho}{\cosh^3\rho},\quad
B^3=\frac\omega{\cosh^3\rho}\end{gather*} and
\begin{gather*}B^1=-\frac{\omega\sin\theta\sinh\rho}{\cosh^4\rho},
\quad B^2=\frac{\omega\cos\theta\sinh\rho}{\cosh^4\rho},\quad
B^3=\frac\omega{\cosh^4\rho}\end{gather*} correspondingly, where we
denote $r=\sinh \rho/\mu$.

Thus we have found the complete set of integrable planar models of neutral particles with Pauli
interaction. The Hamiltonians of these models are given by equation
(\ref{H}) where $\mathbf{B}$ is the vector of external
field whose components are presented in Tables 1 and 2.

\section{Algebras and superalgebras of symmetry operators}

Symmetry operators  presented in any item of Table 1 and Table 2  commute each other. In other words, the presented sets of integrals of motion together
together with the corresponding  Hamiltonians form bases of  Abelian
Lie algebras.

In addition,  integrals of motion collected in Table 2 form interesting
superalgebraic structures. Namely, operators presented in Item 4 satisfy the
following relations:
\begin{gather}\label{SA1}Q_6^2={\cal H},\quad [Q_6,{\cal H}]=0,\\
[Q_1,{\cal H}]=[Q_1,Q_2]=0.\label{SA2}\end{gather} where
\begin{gather}\la{SA11} {\cal H}=H+\left(\mu Q_1+\nu\right)^2+c\end{gather}
and $H$ is the corresponding Hamiltonian (\ref{H}). In other words,
operators $Q_1, \ Q_2$ and $\cal H$ form a basis of the Lie
superalgebra whose  odd and even basis elements are $Q_2$ and $<Q_1,
\cal H>$ correspondingly. Relations (\ref{SA1}) specify $N=1$ SUSY.

Relations (\ref{SA1}) and (\ref{SA2}) can be effectively used to
find eigenvectors and  eigenvalues of the corresponding Hamiltonian (\ref{H}).
 Indeed, the commuting hermitian operators $Q_1,\ Q_2$ and $H$ have common
 eigenvectors. To find eigenvectors for the first order
 differential operators $Q_1$ and $Q_2$ is much more easier than for
 the Hamiltonian which is a differential operator of the second
 order. In addition, relation (\ref{SA1}) makes it possible to find
  eigenvalues of $H$ algebraically starting with eigenvalues for
  $Q_1$ and $Q_2$.

  Operators presented in Item 3 of Table 2  together with  the corresponding Hamiltonian
  (\ref{H}) also form a basis of superalgebra since the following relations are satisfied:
  \begin{gather}\begin{split}&\label{SA3} Q_5^2=\widehat {\cal H},\quad [Q_5,\widehat{\cal H}]=0,\\&
[Q_1,\widehat{\cal H}]=[Q_1,Q_5]=0\end{split}\end{gather} where
\begin{gather}\la{SA31} \widehat{\cal H}=H+\mu Q_1.\end{gather}

In addition to their mutual commutativity  the  symmetry
operators presented in Item 1 of Table 2  satisfy the following quadratic relations
\begin{gather}\label{qr}Q_3^2=H+{2\nu}Q_2+{\nu^2},\\
\label{qr2}\left(Q_3-\frac{1}2\right)^2=\left(Q_2+{\nu}\right)^2.\end{gather}
   Thus the corresponding
  models also admit $N=1$ SUSY
  and can be effectively integrated.

  Integrals of motion presented in Item 2 of Table 2 satisfy the following algebraic relations:
  \begin{gather}\label{AL}Q_4^2=\tilde Q_1^2+2\nu
  \tilde Q_1+\mu^2+\nu^2\end{gather}
  which can be used to find eigenvalues of $Q_2$ using (well known)
  eigenvalues of $Q_1$. Moreover, these integrals of motion  together with operators
  $D=x_1P_1+x_2P_2$, $K={r^2}/2$ and the corresponding Hamiltonian
  $H$ (\ref{H}) form a basis of the five-dimensional Lie
  algebra since the following  commutation relations are satisfied:
  \begin{gather} [H, D]=-2\ri H,\quad [K,D]=2\ri K, \quad [K,H]=\ri
  D\la{CA}\end{gather}
  while all the other commutators are trivial.

  Relations (\ref{CA}) characterize conformal algebra so(1,2), thus
  we deal with a model of conformal quantum mechanics
  ( for definitions see, e.g., \cite{burda}).
  \section{Exact Solutions}
  In this section we use symmetries of the models found above to
  construct their exact solutions. Thanks to the presence of
  arbitrary parameters the number of qualitatively different models
  is too large to be considered in one paper,
  and it is the reason why we
  restrict ourselves only to three particular examples.
  \subsection{Neutron in periodic magnetic field}
    Let us start with the relatively simple Hamiltonian (\ref{H}) whith the components of
  magnetic field $\bf B$ given in Item 1 of Table 2. Since this
  magnetic field depends on one spatial variable, it is
  reasonable to restrict ourselves to
  the one-dimensional eigenvalue problem
\begin{gather}\label{1d1}H\psi\equiv\left(-\frac{\p^2}{\p
y^2}+\mu(\sigma_1\cos y+\sigma_2\sin y
  )+\nu\sigma_3\right)\psi=E\psi\end{gather}
   where we denote $x_1=y$.

Hamiltonian in (\ref{1d1}) includes periodic potential, thus  this
equation is a certain  analog of the Bloch problem for electron.
However, the considered eigenvalue problem is related to a neutral
particle and includes a spin dependent potential.

   Hamiltonian (\ref{1d1}) commutes with  operators $Q_2$ and
   $Q_3$.
Thus we can search for common eigenfunctions for $Q_2,\ Q_3$ and
$H$.

The eigenvalue problem for the first order differential operator
$Q_3$:
\begin{gather}\label{ep}\begin{split}& Q_3\psi_k\equiv \left(\sigma_3(P_1-{\nu})-
{\mu}(\sigma_1\cos( y)+ \sigma_2\sin(
y))\right)\psi=k\psi_k\end{split}\end{gather} is easily solvable.
The general solution for equation (\ref{ep}) is the two component
function
\begin{gather}\label{psi2}\psi_k=\left(\begin{matrix}\varphi_1(y)\\
\varphi_2(y)
\end{matrix}\right)\end{gather}
where
\begin{widetext}
\begin{gather}\begin{split}&\phi_1(y)=\exp\left(\frac{\ri(2\nu+1)y}2\right)
((C_1k_-+C_2\lambda_k)\cos(\lambda_ky)+
(C_1\lambda_k+C_2k_-)\sin(\lambda_ky)),
\\&
\phi_2(y)=-\mu\exp\left(\frac{\ri(2\nu-1)y}2\right)
(C_1\cos(\lambda_ky)+
C_2\sin(\lambda_ky)).\end{split}\label{phi1}\end{gather}
\end{widetext} Here $k_-=k-\frac12,\ \ \lambda_k=\sqrt{k_-^2-\mu^2}$,
and the latest quantity must be real if we ask for solutions whose
norm does not turn to infinity with growing $y$. In other words,
admissible values of $k$ are restricted by the condition
\begin{gather}\label{k}\left(k-\frac12\right)^2>\mu^2.\end{gather}

Let us present the admissible values of $k$ more explicitly. First
we note that up to the unitary transformation
$H\to\sigma_3H\sigma_3$ we can restrict ourselves to $\mu>0$. Then,
in accordance with (\ref{k}) there are two possibilities:
\begin{gather}\label{kv}k\geq \mu+\frac12\quad \texttt{or}\quad
k\leq -\mu+\frac12.\end{gather}

Since operator $Q_2$ commutes with $H$, eigenfunctions (\ref{psi2}),
(\ref{phi1}) solve also equation (\ref{1d1}). The corresponding
eigenvalues $E$ are easily calculated using algebraic relation
$Q_2^2=H$:
\begin{gather}\label{E}E=k^2.\end{gather}

In accordance with (\ref{kv}), (\ref{E})  the admissible values of $E$ are
restricted by the following conditions:
\begin{gather}\label{EE}\begin{split}&E\geq\left(\mu-\frac12\right)^2\quad
\texttt{if}\quad \mu>\frac12,\quad k<\frac12-\mu,\\& E\geq 0 \quad
\texttt{if}\quad 0<\mu\leq\frac12,\ \ k\leq
\frac12-\mu,\\&E\geq\left(\mu+\frac12\right)^2\quad \texttt{if}\quad
 k\geq\mu+\frac12.\end{split}\end{gather}

 The probability
density corresponding to solutions (\ref{phi1}), i.e.,
\begin{gather}\label{pd}\begin{split}&\phi_1\phi_1^*+\phi_2\phi_2^*=(C_1^2+C_2^2)k_-^2
+2k_-\lambda_kC_1C_2+\frac14(C_1^2-C_2^2)\mu^2\cos(2\lambda_ky)\end{split}\end{gather}
is a periodic function. However its period can differ from the shift
which keeps equation (\ref{1d1}) invariant, i.e., from $2\pi $. Such
situation looks rather nonphysically, since the probability density
calculated in a fixed frame of reference can differ from the density
calculated in the equivalent frame of references shifted by $2\pi $,
in spite of that the equation (\ref{1d1}) in these frames has
exactly the same form.

 There are two ways to obtain solutions whose
amplitude is a periodic function with the period $2\pi$. First it is
possible to choose in (\ref{phi1})
\begin{gather}C_1=\pm C_2=\frac1{2\sqrt{\pi k_-(k_-\pm\lambda_k)}}\end{gather}
 and obtain solutions normalized at any invariance interval
 $[y,y+2\pi]$:
\begin{gather*}\begin{split}&\phi_1(y)=
\frac12\sqrt{\frac{k_-\pm\lambda_k}{\pi k_-}}
\exp\left(\frac{\ri(1-2\nu\pm2\lambda_k)y}2\right),\\&
\phi_2(y)=\frac\mu{2\sqrt{\pi
k_-(k_-\pm\lambda_k)}}\exp\left(-\frac{\ri(2\nu+1\mp2\lambda_k)y}2\right)
.\end{split}\end{gather*}
In this case there are no restrictions on eigenvalues $E$  additional to (\ref{EE}).

The second way is to impose the following condition on the spectral
parameter $k$:
\begin{gather}\label{spectr}k=\frac12(\varepsilon\sqrt{n^2+4\mu^2}+1)
\end{gather} where $n=0,1,2,\dots,\quad \varepsilon=\pm1.$ In this case the
energy levels (\ref{E}) are discrete  and there are two branches:
$E_+$ and $E_-$ where
\begin{gather}\la{E1}
E_\pm=\frac14(n^2+4\mu^2\pm2\sqrt{n^2+4\mu^2}+1).
\end{gather}  The corresponding
eigenfunctions can be obtained from (\ref{psi2}), (\ref{phi1})
changing $\lambda_k\to\frac{n}2 $ and using expression
(\ref{spectr}) for $k$. To obtain normalized solutions, arbitrary
constants $C_1$ and $C_2$ should be restricted by the following
condition:
\[(C_1^2+C_2^2)(n^2+4\mu^2)+2\varepsilon
nC_1C_2\sqrt{n^2+4\mu^2}=\frac2\pi.\]

Thus like in the Bloch problem for electron \cite{bloch} the
energies of neutron moving in the periodic field represented in Item 1 of Table 2 can be
continuous and have a band structure, see (\ref{EE}). In addition,
there are solutions with discrete spectrum (\ref{E1}).

\subsection{Rotationally invariant system}The next system which we
consider includes the following Hamiltonian:
\begin{gather}\label{2d1}H=-\nabla^2+\frac\mu{r^3}(\sigma_1x_2-\sigma_2x_1)
+\frac{\alpha}{r}\end{gather}where $\mu$ and $\alpha$ are real
parameters.

If $\alpha=0$ Hamiltonian (\ref{2d1}) coincides with operator
(\ref{H}) where $B^1$ and $B^2$ are components of the external field
presented in Item 2 of Table 2 with $k=1$ and $\nu=0$. For nonzero
$\alpha$ this is a Hamiltonian of type (\ref{HaHa}).

The additional term $\frac{\alpha}{x}$ which we include to obtain a
more general model does not break the commutativity of the
Hamiltonian with operators $Q_1$ and $Q_4$. Using this fact and taking
into account relations (\ref{AL}) we can expand solutions of the
eigenvalue problem (\ref{EP}) for Hamiltonian (\ref{2d1}) via
eigenvectors of these operators  satisfying
\begin{gather}\label{2d2}\begin{split}&\left(L+\frac{\sigma_3}2\right)
\psi_{k,\varepsilon} =k\psi_{k,\varepsilon},\quad k=\pm\frac12, \
\pm\frac32, \dots,
\\&\begin{split}&
\left(\sigma_3L-\frac\mu{r}(\sigma_1x_2-\sigma_2x_1)+
\frac12\right)\psi_{k,\varepsilon}=
\varepsilon\sqrt{k^2+\mu^2}\psi_{k,\varepsilon},\quad
\varepsilon=\pm1.\end{split}\end{split}\end{gather}

Solutions of equations (\ref{2d2}) can be represented in the
following form
\begin{gather}\label{2d4}
\psi_{k,\varepsilon}=\frac{C_{k\varepsilon}\phi(r)}{\sqrt{r}}\left(\begin{matrix}\exp(i(k-\frac12)\theta)
\left(k+\varepsilon\sqrt{k^2+\mu^2}\right)\\
\ri\exp(i(k+\frac12)\theta)\mu\end{matrix}\right)\end{gather} where
$C_{k\varepsilon}$ are integration constants and polar coordinates
(\ref{pc}) are used. Substituting (\ref{2d1}), (\ref{2d4}) and
(\ref{pc}) into (\ref{EP}) we obtain the following ordinary
differential equation for radial function $\phi=\phi(r)$:
\begin{gather}\left(-\frac{\p^2}{\p
r^2}+\frac{k^2-\varepsilon\sqrt{k^2+\mu^2}}{r^2}-
\frac{\alpha}r\right)\phi=E\phi.\label{2d5}
\end{gather}

Let $\alpha=0$ then equation (\ref{2d5}) with the boundary condition
\begin{gather}\phi=0 \quad \texttt{if}\quad r=0\la{IC}\end{gather}
defines the eigenvalue problem of one dimensional conformal quantum
mechanics. Its solutions can be expressed as follows:
\begin{gather}\label{2d6}E=p^2>0,\quad \phi =\sqrt{r}J_\nu(pr)\end{gather} where $J_\nu(pr)$ is the
Bessel function with
\begin{gather}\la{nu}\nu=\frac12\sqrt{1+4k^2-
4\varepsilon\sqrt{k^2+\mu^2}}.\end{gather} If $\varepsilon=1$ then admissible
values of parameters $\mu$ and $k$ are constrained by the following
relation:
\begin{gather}\la{2d7}\left(k^2-\frac14\right)^2\geq\mu^2 \quad
\texttt{if}\quad \varepsilon=1.\end{gather}

 Let $\alpha>0$ then solutions of
equation (\ref{2d5}) can be found using its analogy with the radial
equation for Hydrogen atom.  The eigenvalues $E$ which correspond to
square integrable solutions vanishing at the singularity point $r=0$
look as follows:
\begin{gather}E=-\frac{\alpha^2}{4\left(n+\frac34+\nu\right)^2}\label{EEE}\end{gather}
where $\nu$ is parameter given in (\ref{nu}), $n$ is a natural
number and $k$ satisfies conditions (\ref{2d2}) and (\ref{2d7}). The
corresponding eigenvector $\phi$ can be expressed via a linear
combination of Whittaker functions $M(a,b,x)$ and $W(a,b,x)$:
\begin{gather} \phi=C_1M\left(a,b,\frac{\alpha
y}{a}\right)+C_2W\left(a,b,\frac{\alpha
y}{a}\right)\la{ev}\end{gather} where
\[a=n+\nu+\frac34,\quad b=\nu+\frac14.\]

Notice that in contrast with the Hydrogen atom there is not a
degeneration w.r.t. orbital quantum number. However, any energy
level is infinitely degenerated since the corresponding eigenvector
(\ref{ev}) includes two integration constants, an only one of them
can be fixed by normalizing the wave function.

\subsection{Shape invariant system}
  Let us consider the eigenvalue problem for Hamiltonian (\ref{Ha})
  where $\bf B$ is the magnetic field whose components are given in Item 4 of Table 1 where $f_3=0$:
    \begin{gather}\begin{split}&H\psi\equiv\left(-\nabla^2+\lambda(1-2\kappa)\exp(-x_2)(\sigma_1\cos
    x_1-\sigma_2\sin
    x_1)+\lambda^2\exp(-2x_2)\right)\psi= E\psi.\end{split}
    \label{EP1}\end{gather}
    Here $\lambda$ is the integrated coupling constant and new parameter
    $\kappa$ is introduces such that
     $\mu=\lambda(1-2\kappa)$.

Hamiltonian $H$ in (\ref{EP1}) admits integral of motion $Q_2=P_1-\frac{\sigma_3}2$. Thus it is possible to expand solutions of
(\ref{EP1}) via eigenvectors of $Q_2$ which look as follows:
\begin{gather}\label{psi}\psi_p=\left(\begin{array}{cc}
\exp(i(p+\frac12)x_1)\varphi(x_2)\\\exp(i(p-\frac12)x_1)
\xi(x_2)\end{array}\right)\end{gather} and satisfy the condition
$Q\psi_p=p\psi_p$.

Substituting (\ref{psi}) into (\ref{EP1}) we come to the following
equation:
\begin{gather}\label{equ1}\left(-\frac{\p^2}{\p
y^2}+V_\kappa\right)\Phi=\varepsilon\Phi\end{gather} where we denote
\begin{gather}\label{defi}y=x_2,\quad \varepsilon=E-p^2-\frac14,\quad
\Phi=\left(\begin{matrix}\varphi\\
\xi\end{matrix}\right)\end{gather} and
\begin{gather}\label{V}
V_\kappa=\lambda^2\exp(-2y)-\lambda(2\kappa-1)\exp(-y)\sigma_1-p\sigma_3.\end{gather}
We can restrict ourselves to non-positive $p$, then solutions for
$p>0$ could be obtained changing $\Phi\to\sigma_1\Phi$.

Let us consider the eigenvalue problem (\ref{equ1}) with the
following conditions:
\[y\geq0,\quad \Phi(0)=0,\quad \int_0^\infty
\Phi^*(y)\Phi(y)dy<\infty.\]

 Potential
(\ref{V}) belongs to the list of shape invariant matrix potentials
found in \cite{NK1}, see equation (5.12) therein. It can be
represented in the form $V=W_\kappa^2-W_\kappa'+c_\kappa$ where
\begin{gather}\label{exp}\begin{split}& W_{\kappa}= -\kappa+ \lambda\exp(-
y)\sigma_1-\frac{p}{2\kappa}\sigma_3,\qquad c_\kappa=\kappa^2+\frac
{p^2}{4\kappa^2}.\end{split}\end{gather}

The shape invariance means \cite{Gen} that the superpartner
potential
    $V_\kappa^+=W_\kappa^2+W_\kappa'+c_\kappa$ is equal to the
    initial potential with shifted parameter $\kappa$
    up to a constant term. And it is the case for potential (\ref{V}) since
\begin{gather*}W_\kappa^2+W_\kappa'=
W_{\kappa+1}^2-W_{\kappa+1}'+c_\kappa-c_{\kappa+1}.\end{gather*}

Using the shape invariance it is possible to integrate equation
(\ref{EP1}) in a simple and straightforward way with using tools of
SUSY quantum mechanics \cite{NK1}. The eigenvalues $\varepsilon$ and
the corresponding state vectors are enumerated by natural numbers
$n=0,1,\dots$ The ground state vector $\Phi_0(\kappa,y)=
\left(\begin{matrix}\varphi_0\\
\xi_0\end{matrix}\right)$ should solve the equation
\begin{gather}
  \label{psi0}
  a_{\kappa}^-\Phi_0(\kappa,y)\equiv
  \left( \frac{\p}{\p
  x}+W_{\kappa}\right)\Phi_0(\kappa,y)=0,
  \end{gather}
thus
\begin{gather}\label{GS1}\varphi_0=z^{\frac12-\kappa}
K_{\nu+1}(z), \quad \xi_0=z^{\frac12-\kappa} K_{\nu}(z)
\end{gather} where $K_\nu(z)$ is modified Bessel function, $\nu=\frac{p}{2\kappa}-\frac12$ and
 $z=\lambda\exp(-y)$.

 Solutions
  which correspond to $n^{th}$ exited state
  can be calculated using the following relation:
\begin{gather}\label{psin}\Phi_n(\kappa,y)=
a_{\kappa}^+a_{\kappa+1}^+ \cdots a_{\kappa+n-1}^+\Phi_0(\kappa+n,y)
\end{gather}
where $a_\kappa^+=-\frac{\p}{\p y}+W_\kappa(y)$. Finally,
  the corresponding values of spectral parameter $\varepsilon$ have the
  following form
   \cite{NK1}
\begin{gather}\label{EV1}\varepsilon=-N^2-\frac{p^2}
{4N^2}\Rightarrow E=p^2-N^2-\frac{p^2} {4N^2}+\frac14\end{gather}
where $N=\kappa+n$ and $n$ is a natural number.

In accordance with (\ref{EV1}) eigenvalues $E$ are invariant w.r.t.
the change $p\to-p$, thus eigenvectors corresponding to a chosen $n$
are linear combinations of functions (\ref{psi}):
\begin{gather}\la{lc}\begin{split}&\Psi_n=C_1\left(\begin{matrix}
\exp(i(p+\frac12)x_1)\varphi_n(x_2)\\\exp(i(p-\frac12)x_1)
\xi_n(x_2)\end{matrix}\right)+C_2\left(\begin{matrix}
\exp(i(-p+\frac12)x_1)\xi_n(x_2)\\\exp(i(-p-\frac12)x_1)
\varphi_n(x_2)\end{matrix}\right).\end{split}\end{gather}

Let $C_1C_2\neq0$ then, in order the norm of this function be
invariant w.r.t. the shifts $x_1\to x_1+2\pi$ like the Hamiltonian
(\ref{EP1}), it is necessary to impose the following condition:
\[p=\frac{2m+1}2,\quad m=0, 1, \dots\]  For solutions
(\ref{lc}) with $C_1\equiv0$ or $C_2\equiv0$ the spectral parameter
$p$ can be quantized by imposing the periodic boundary  condition
with an arbitrary period.

\section{Discussion}

In the present paper planar Schr\"odinger-Pauli equations for
neutral particles, which admit first order constants of motion, are
classified. The collection of such equations appears to be rather
rich and interesting. In particular, it includes supersymmetric
systems belonging to $N=1$ SUSY quantum mechanics, the system with
shape invariant Hamiltonian (\ref{EP1}) and the system with
Hamiltonian (\ref{2d1}) (were $\alpha=0$), which is conformally
invariant.

Any Hamiltonian (\ref{H}) with the external fields presented in Table 2 admits two integrals of motion, and following \cite{wint6} -\cite{wint8} we can call the related 2d systems superintegrable. Notice that this terminology is rather conventional, since the additional (spin) degree of freedom is ignored. To fix this degree of freedom we need an additional involutive integral of motion like matrix $\sigma_3$, which can extend the possible number of symmetries.  Maybe  it is more natural to say that a 2d system with spin 1/2 is "superintegrable" (integrable) if it has at least three (two) independent integrals of motion. These speculations can be justified by the example given  in Item 6 of Table 1. The corresponding 2d system admits a constant of motion, but it is not convenient to call it "integrable" since the related Hamiltonian is a direct sum of two 2d "nonintegrable"  Hamiltonians which do not admit constants of motion  provided $B_3$ is an arbitrary function of $x_1$ and $x_2$.

Quantum mechanical systems with a sufficiently large number of constants of motion are usually exactly solvable, and it is the
case for the models classified in the above. We restrict ourselves
to solving three of them in section 7. To solve the model whose
Hamiltonian is given by equation (\ref{EP1}) we use its shape
invariance, and this is the second direct application of results of
paper \cite{NK1} where matrix superpotentials were classified, to a
d-dimension models with d$>1$. The first application of these
results to planar systems with arbitrary spin can be found in paper \cite{N2}).

A physically interesting subclass of the classified systems includes
Hamiltonians (\ref{1d1}) and (\ref{EP1}) whose effective potentials
are periodic functions. Such potentials simulate interaction of
neutron with a crystal lattice.

Fundamental results concerning the motion of electron in periodic
electromagnetic field was formulated long time ago by F. Bloch
\cite{bloch}. Our analysis of solutions (\ref{phi1}) shows that  the neutron interacting with a
periodic magnetic field can have both continuous and discrete energy
spectrum.

The presented classification of  planar Hamiltonians (\ref{H}) admitting first order integrals of motion is
complete. Nevertheless it can be considered as an intermediate
outcome. First it is interesting to study integrable planar
systems with higher order integrals of motion. An example of such
system is the Pronko-Stroganov model \cite{pron1} which admits
symmetry operators of second order. Then, our analysis can be
extended to systems of type (\ref{Ha}) with arbitrary $\bf B$ and
$V$.  Finally, the 3d superintegrable systems for neutrons are also
waiting for their classification. In other words, superintegrable
systems with spin whose investigation was started with paper
\cite{wint6} belong to a promising research field.

\begin{acknowledgments}
I am indebted to Prof. Petr Reimer for his kind invitation to visit the Division of Elementary Particle Physics of Institute of Physics, Nat. Acad. Sci. of Czech Republic, were  this work had been  finished.
\end{acknowledgments}

\end{document}